\documentstyle[preprint,aps]{revtex}
\begin{document}
\draft
\preprint{}

\title{2D-3D Vortex Line Transition in Two-Layer Josephson Arrays}

\author{Wenbin Yu and D. Stroud\\
	{\it Department of Physics, The Ohio State University,
	Columbus, OH 43210}}

\date{\today}
\maketitle

\begin{abstract}

We calculate the dynamical response of two-layer Josephson arrays
to a bias current applied into only one layer 
(primary layer), in a perpendicular magnetic field.
The pancake vortices in the two layers 
sometimes form into
flux lines which move as a unit under the influence of the bias current.
The lines break apart into independent pancakes in sufficiently high
currents, sufficiently weak interlayer coupling, or possibly
sufficiently high fields.  
We discuss the relevance of these results to recent experiments in 
high-T$_c$ superconductors in the flux transformer geometry.

\end{abstract}



\maketitle

\newpage

\section{Introduction}

The so-called ``flux-transformer'' geometry\cite{Giaever,Ekin}. 
provides a novel way to study vortex order in Type-II superconductors.
In this technique, the external
current density ${\bf J}$ is injected into and parallel to 
the upper $ab$ plane of a
high-T$_c$ superconductor.  The voltage drop is
measured across both the upper plane (the ``primary voltage'') and the
lower plane (the ``secondary voltage'').
For applied magnetic field ${\bf B} \| c$,
the geometry is sensitive to the vortex order, especially to the
interlayer coupling between the so-called vortex ``pancakes'' in the 2D 
layers.  Typically, this coupling will cause
the pancakes to form into flux lines.  
The ${\bf J} \times {\bf B}$ Magnus
force will exert a torque on such a flux line.  If the interlayer coupling
is strong enough, the flux line will move as a unit under this force,
producing equal primary and secondary voltages.
But at sufficiently large ${\bf J}$, for any large anisotropy, the 
influence of
the torque will exceed that of the interlayer coupling, and
the flux line will break.  Even at small
applied currents, the flux line can sometimes be broken
by thermal noise.
In both cases, the line breaking shows up
in the difference between the primary and the secondary voltages, which 
increases quickly from zero as the pancakes decouple.)
This cross-over from 3D flux lines to 2D vortex pancakes has attracted great 
experimental and theoretical 
interest\cite{Clem,Safar,Fisher,Nelson,White,Glazman,wan,Lopez,Cruz}.

In this paper, rather than treating the actual high-T$_c$ superconductor
directly, we model the $T=0$ flux line dynamics in the flux-transformer 
geometry, using a two-layer Josephson junction array.  While 
far from a realistic high-T$_c$ material, this model is numerically 
tractable\cite{teitel,lee}, 
and in addition could be directly studied 
experimentally.   We find that in this flux-transformer
geometry, the vortex lines do exhibit a 3D-2D decoupling transition at
a critical vortex velocity,
just as in the high-T$_c$ materials.  We map out this transition as a 
function of anisotropy and applied current, and find behavior
which qualitatively resembles experiment.
 Our calculations show
that this transition can be caused by
Josephson coupling alone, without other types of interactions
(e.\ g., electromagnetic interactions\cite{Giaever,Ekin},  Coulomb
interactions\cite{Duan}) between the pancakes on different layers.

\section{Model}

We consider a two-layer square arrangements of Josephson 
junctions (cf.\ Fig.\ 1).
All the junctions, both 
within and between the planes, are assumed to be overdamped, resistively 
shunted junctions (RSJ model).
All intralayer and interlayer junctions are assumed identical, with 
critical currents and resistances  $I_c, R$ and $I_{c^{\prime}}, R$ 
respectively.  The ratio $\alpha \equiv I_{c^{\prime}}/I_c$ represents the 
array anisotropy, and is the analog of the high-T$_c$ mass anisotropy.

To simulate the flux-transformer geometry, we inject a bias
current along the $y$ direction into only the upper layer 
(the ``primary'' layer; cf.\ Fig.\ 1)\cite{wan}.  We denote the other 
the {\em secondary} layer.
To introduce vortices, we apply a magnetic field
${\bf B} \| z$, i.~e., perpendicular to the layers.
The Magnus (${\bf J} \times {\bf B}$) force from the transport current 
will drive these vortices in the $x$ direction.

The equations of motion for this system take the standard form
$I_{ij} = V_{ij}/R_{ij} + I_{c;ij}\sin(\phi_i - \phi_j - A_{ij}$, 
$V_{ij} \equiv  V_i - V_j = (\hbar/2e)(d/dt)(\phi_i - \phi_j)$, and
$\sum_jI_{ij} = I_{i;ext}$.
Here $I_{ij}$ is the total current from grain $i$ to grain $j$; $R_{ij}$ and 
$I_{c;ij}$ are the shunt resistance and critical current of the Josephson 
coupling between grain $i$ and grain $j$; $\phi_i$ is the phase
of the order parameter on grain $i$; and $I_{i;ext}$ is the external 
current fed into grain $i$.  $V_{ij}$ and $A_{ij}$ are the voltage
difference and magnetic gauge phase factor between grain $i$ and grain $j$.
$A_{ij}=(2\pi/\Phi_0)\int_{\bf{x}_i}^{\bf{x}_j}\bf{A}\cdot\bf{dl}$,
where $\Phi_0 \equiv hc/(2e)$ is the flux quantum, $\bf{A}$
is the vector potential for the applied magnetic field and $\bf{x}_i$ is the 
position of the center of grain $i$.  
The first equation describes the total current between grains $i$ and $j$ 
as the sum of two terms: an Ohmic current through the shunt resistance, and 
a Josephson current.  The second is Josephson equation relating voltage 
and phase, while the last is just Kirchhoff's Law, describing current 
conservation at the $i^{th}$ node.

As boundary conditions, we introduce a uniform current 
$I_{i;ext}=I$ into each boundary grain in the primary layer and extract
it from the other edge (cf.\ Fig.\ 1). 
In the secondary layer, we use free boundary
conditions in the $y$ direction. In both layers, periodic boundary 
conditions are adopted in the $x$ direction.
$A_{ij}=0$ for an interplanar junction, 
while for an intraplanar junction 
$\sum_{plaquette} A_{ij} = 2\pi \frac{BS}{\Phi_0} \equiv 2\pi f$ 
($S$ is the plaquette area)\cite{yu.1}.
We consider only $f=1/N^2$ so that 
there is only one vortex (i.\ e.\ one flux quantum) piercing each layer. 
We denote these the primary and secondary vortices. 

We solve the coupled equations of motion
as described previously\cite{yu.1}, using a fourth-order 
Runge-Kutta algorithm. During the calculation of the dynamical
state $\{\phi_i(t)\}$, we keep track of the vortex motion in the
array, in terms of the vortex number $n$ of each plaquette.  $n$ is 
defined by
$\sum_{plaquette} (\phi_i-\phi_j-A_{ij}) + 2\pi f = 2\pi n$,
where 
$(\phi_i-\phi_j-A_{ij})$ is in the range $(-\pi,\pi]$ and the 
summation is taken in the counterclockwise direction.

\section{$I-\langle V\rangle$ Characteristics}

Our $I-\langle V\rangle$ characteristics typically fall
into one of five regimes. (i) At the lowest currents,
$\langle V\rangle = 0$:
both primary and secondary vortices are pinned by the
periodic potential of the array\cite{lat,rzch}. (ii) At slightly higher
currents, in some cases, the primary vortices move, 
while the secondary vortices are still pinned. 
(iii) In other arrays,
at such currents, both primary and secondary vortices are depinned and move
together as a single flux line. (iv) At still higher currents, in some
arrays, both vortices are depinned, but move 
{\em independently}, as separate ``pancakes''. (v) For $I > I_c$,
the entire Josephson lattice is depinned, and the $I-\langle V\rangle$ 
characteristics are dominated by single-junction effects. 

Fig.\ 2 shows the calculated $I-\langle V\rangle$
characteristics of a $2 \times 10 \times 10$ ($N=10$) array at 
$f=0.01$ and three
values of $\alpha$. In the flux-flow regime [cases (ii)-(iv)], 
the $I-\langle V\rangle$ curves are dominated by vortex motion.
In all cases, whether moving or fixed, these
vortices remain in the center of the array (i.~e.\ the 5$^{th}$ row),
presumably the region of lowest energy. (In our $19\times 19$ array
calculations, they stay in the $10^{th}$ row.)

The critical current for the onset of voltage is very sensitive to the
anisotropy $\alpha$.  When $\alpha = 0$, Fig.~2(b) shows that
the primary vortex is depinned at the bias current $I_{cp} \approx 0.1I_c$,
while the secondary vortex remains fixed up to a bias current
$I_{cs} \approx 0.28I_c$.  At $\alpha = 0.01$, these
depinning currents are respectively $0.16I_c$ and $0.26I_c$.
But when $\alpha = 0.04$, the primary and the secondary vortices
are depinned {\em simultaneously} near $0.17I_c$, and move
thereafter as a single flux line in the array.  This line, however, breaks
apart near $I_{cd} = 0.36I_c$, above which the
primary vortex moves faster than the secondary vortex.
For all three $\alpha$ values, the individual junctions in the
primary layer go normal near $I_{cj} = 1.06I_c$. 

In the zero-coupling limit, the value $I_{cp}=0.1I_c$
[cf.\ Fig.~2(b)] is the depinning current of a single
vortex in a 2D array\cite{lat}.
This can be understood by the following argument.   Below $I_{cp}$, 
there is zero voltage drop across
every junction.  Hence, no current passes through the resistive shunts
between layers.  Moreover, since $\alpha = 0$, there is no interlayer 
Josephson current.  Therefore, all the bias current passes only
through the primary layer, and $I_{cp}$ must be the 
vortex depinning current for a single-layer array.  
Once the primary vortex has been depinned, there is
some voltage drop between the layers.  This voltage drop produces currents
into the secondary layer via the the interlayer shunt resistances.
For sufficiently large $I$, these
currents generate a Magnus force
which depins the secondary vortex.

At sufficiently small {\em nonzero} $\alpha$, we still find that
$I_{cp} < I_{cs}$.   But above a critical $\alpha_c$, $I_{cp} = I_{cs}$, and
the two vortices move together as a single flux line for $I > I_{cp}$.
$\alpha_c$ seems also to be size-dependent.

We can crudely understand all of these phenomena from a simple argument. The
idea is illustrated schematically in Fig.~3, which shows a side
view of an $N = 11$ array with transport current
injected from the left.  The primary and secondary vortices -- both located in
the middle row of the array --  are indicated by crosses.  Since the vertical
junctions have critical current $\alpha I_c$, the total current passing through
these junctions into the secondary layer, under zero voltage conditions, 
cannot exceed $\approx \frac{1}{2}\alpha NI_c$
to the left of the vortex.  Assume that the transport current
injected into the primary layer will divide equally between the two layers
until this limit is exceeded.  This occurs at a transport current
$I_{eq} = \alpha N I_c$.
From this picture, we can estimate the conditions on $N$ and $\alpha$
such that the two vortices will be depinned {\em simultaneously}.
Namely, if $N\alpha/2 > 0.1$, then {\em both} vortices will be depinned
simultaneously at current below this limiting current and move together
thereafter under the bias current drive.
Conversely, if $N\alpha/2 \leq 0.1$, the vortices will be depinned at
different currents, and will move more or less independently.
(Note that if there were no resistive shunt coupling between layers, 
then at $\alpha =0$ value,
the two layers would be entirely independent
the secondary vortex would never be depinned.)

For $N=10$, the above argument gives $0.02 < \alpha_c < 0.03$,
consistent with our numerical results. For an $N=19$ array,
the corresponding value of 
$\alpha_c$, both analytically and numerically, 
is in the range $0.01<\alpha_c <0.02$. To test this picture
further, we have also calculated the current distribution in the array for a
variety of parameters.  The distribution of currents in the vertical junctions
indeed conforms with the estimates given above.

This picture may possibly explain the breaking of the flux line at large 
enough $I$.  As $I$ increases, the line moves faster and faster, but the 
secondary vortex will lag the primary one by an increasing distance.
Eventually, this distance may be such that (in a small array) the secondary 
vortex will start to feel the influence of the {\em images} of the 
primary vortex.  Above this current, the primary and secondary vortices
should break apart and move separately.  

These arguments suggest two conclusions.
First, for large enough $N$, the primary and the secondary vortex 
will {\em always} form a single flux line -- that is, will 
be depinned and move together --
for an arbitrarily weak but nonzero interlayer Josephson coupling $\alpha$.
Secondly, in a sufficiently large array, 
the line should remain unbroken all the way up to the single-junction 
critical current $I_c$\cite{comment}.  We have carried out a few 
calculations for larger arrays, which do support these conclusions.

We can also speculate how the 3D/2D transition would depend on magnetic 
field strength.  Since $f = 1/N^2$ in our simulations, decreasing the array 
size may be comparable to decreasing the field strength.
Our simulations at different $N$, combined with the above arguments,
suggest that
$\alpha_c$ \underline{increases} with increasing $f$  --
that is, that flux lines are more easily broken in high flux density
superconductors.  Hence, for a given weak interlayer coupling, 
there may be an upper limiting field,
above which there are no 3D flux lines at all, but only 2D pancake
vortices.  Such an upper limit may have
been observed in experiments\cite{White} and has been predicted 
theoretically\cite{Fisher,Glazman}.

\section{Time-Dependent Behavior}

Fig.\ 4(a) shows both $V_p(t)$ and $V_s(t)$ for $N = 10$,
$I = 0.2 I_c$, and $\alpha = 0$. 
For this case, we find that
the primary vortex moves across the array in the
$5^{th}$ row, while the
secondary vortex remains fixed in the plaquette (10, 5).
Correspondingly, $\langle V_p\rangle \neq 0$ but $\langle V_s\rangle = 0$.
(Similar results are found for larger arrays.)
The peaks in $V_p(t)$
correspond to the translation of the primary vortex by one plaquette.
The corresponding peaks in $V_s(t)$,
resulting from the weak resistive coupling between the layers,
correspond to oscillations of the secondary vortex around an equilibrium 
position.  Since the interlayer resistive coupling is very weak, there is a
constant time interval between 
consecutive peaks in $V_p(t)$, implying that the velocity of the
primary vortex does not depend on its proximity to
the secondary vortex, and suggesting that the
intervortex interaction nearly vanishes under these 
conditions. The time interval $T$ between peaks 
satisfies the Josephson relation
$\langle V\rangle = \frac{\hbar}{2e}\frac{2\pi}{NT}$, where 
$\langle V_p \rangle$ is the time-averaged voltage difference across the 
primary layer.  This is consistent with the picture of a vortex circling 
the array once in an interval $NT$.

Fig.\ 4(b) shows the corresponding voltages at $N=10$, $I=0.16I_c$, and
$\alpha =0.01$. In this case, the primary vortex is depinned, traveling in the
$5^{th}$ row of the array, while the secondary vortex remains fixed in 
plaquette (10, 5),  as in the $\alpha = 0$ case. 
Once again, both $V_p(t)$ and $V_s(t)$ exhibit many peaks, with the same 
interpretation as in 4(a).  But now the interval between peaks on $V_p(t)$
is variable, implying a plaquette-dependent 
primary vortex velocity.  The 
interval is smallest, and the average
vortex velocity greatest, when primary and 
secondary vortices interval are directly superposed, and the
mutual attraction is strongest.  Because of the periodic boundary 
conditions, $V_p(t)$ is periodic, with $N$ peaks in each 
period.  The peak structure in $V_p(t)$ proves that an attractive 
interaction between vortices can be produced by interlayer Josephson 
coupling alone.

Fig.~5 shows $V(t)$ and the vortex paths for $N = 10$, 
$f=0.01$, $\alpha =0.04$, and $I=0.18I_c$. In this case,
the two pancake vortices form a single flux line which moves as a unit at low
bias current. Both $V_p(t)$ and $V_s(t)$ are characterized by periodic
double-peaked structure [cf.\ Fig.~5(a)], which can be understood
from the vortex spatial path shown in Fig.~5(b). The larger
peaks of $V_p(t)$ and the smaller peaks of $V_s(t)$ correspond to the 
translation of the primary vortex by one plaquette,  while the other peaks
arise from the translation of the secondary vortex.  
$V_p(t)$ and $V_s(t)$ have the same frequency.
Although the two vortices move as a unit, the vortex line is
\underline{at an angle} to the $c$ axis: the primary vortex leads the secondary
vortex on average by about half a plaquette.
In simulations at $N = 19$, we have found
that the primary vortex can lead the secondary vortex by as much as 
two plaquettes.  As $\alpha$ increases, this horizontal distance decreases, 
showing that the interlayer vortex interaction is increasing.

Note also that $V_p(t)$ and $V_s(t)$ have slightly different waveforms, 
although they have the same frequency. This means that the flux line formed 
by the two pancakes is not rigid. 
Our picture is that the flux line moves as a unit
across the array in a periodic motion, but that the two constituent 
pancakes may oscillate around their steady state positions on the flux
line. The period of such an
oscillation would equal that of the flux line
motion. 

At $I=0.36I_c$, for this coupling, the flux line breaks into
independent primary and secondary pancakes.   Above this
current, $\langle V_p\rangle > \langle V_s \rangle$, indicating
that the primary vortex moves faster than the secondary
one.  Such a 
critical vortex velocity for flux line breaking, i.~e., a 3D to 2D crossover, 
has been reported in high-T$_c$ superconductors\cite{Cruz}.

\section{Discussion and Conclusions}

We have studied the dynamical response of
two-layer Josephson arrays to a bias current applied in the flux 
transformer geometry.  Although this model is far from a realistic high-T$_c$
superconductor, our numerical results agree qualitatively with some 
experimental studies on such materials.  Thus, they may be useful in 
constructing more realistic models for the dynamics of high-T$_c$ materials 
in this geometry.

Our results show that interlayer Josephson coupling alone is enough to bind
pancake vortices into vortex lines. 
These lines move as a unit under the drive of an external bias current,
but may break up into 2D pancake 
vortices and move independently at high enough currents,
provided the magnetic field is not too small.  Our calculations
also suggest that, for any given current, there may exist an upper 
magnetic field limit above which 3D flux lines are unstable in the flux 
transformer geometry.

Our results suggest several areas for future work.  First, it would be of 
interest to extend this work to larger magnetic fields and finite
temperatures (included via Langevin 
noise), in order to study the 3D-2D transition at finite fields and
temperatures.  Likewise, the inclusion of a finite inductive coupling
between plaquettes\cite{dominguez,vanderzant} might lead to novel
behavior not seen in the present, low-screening limit.  Finally, the 
predicted effects might be sought experimentally, not only in
high-T$_c$ materials, but directly in two-layer Josephson arrays which 
could be made in the geometries described here.

\section{Acknowledgements.}

We thank Prof.\ S. M. Girvin and  Dr.\ Saad Hebboul for drawing this 
problem to our attention, and Dr.\ Y. M. Wan for valuable conversations.
This work was supported by NSF grants DMR 90-20994 and
DMR94-02131, and by the Midwest Superconductivity Consortium at Purdue
University through D. O. E. Grant DE-FG90-02-ER-45427. 
Calculations were carried out, in part, on the CRAY Y-MP 8/864 of the 
Ohio Supercomputer Center.

\newpage
\begin{center}
FIGURE CAPTIONS
\end{center}
\vspace{0.1in}

\begin{enumerate}

\item
Schematic diagram of a $2 \times 5\times 5$ ($N = 5$) two-layer Josephson
junction array.  Each intersection denotes a superconducting grain.
Each grain is coupled to its four intralayer neighbors, and to the nearest 
grain in the adjacent layer, by a resistively 
shunted Josephson junction (RSJ).
An external magnetic field ${\bf B}$ is applied
perpendicular to the layers in the $z$ direction. A bias current
$I$ is injected into each grain at one edge of the upper layer (the
``primary layer''), and extracted at the opposite
edge in the $y$ direction. 
$V_p$ and $V_s$ denote the voltage drops across the primary and secondary
layers (averaged over $x$), as shown.  There are periodic boundary 
conditions in the $x$ direction.

\item
$I-\langle V\rangle$ characteristics for an $N = 10$
at $f=0.01$, and three different values of $\alpha$.
In each case, the two curves denote $V_p$ and $V_s$.
(a) $\alpha= 0$; (b) $\alpha=0.01$; (c) $\alpha=0.04$.

\item
Schematic view of the array, viewed along the $x$ direction, and
showing the current direction in junctions
parallel to $y$ axis in an $N = 11$ array.
Arrows denote the direction of current flow; crosses denote
the vortex pancake positions.

\item
$V_p(t)$ (full curves) and $V_s(t)$ (dotted curves) in $N = 10$ 
arrays at $f=0.01$ for two coupling ratios and bias
currents as indicated.

\item
Time-dependent voltage traces and vortex path for an
$N=10$ two-layer array at $f=0.01$, $\alpha =0.04$,
and $I=0.18I_c$.  In both cases, the full curve corresponds to the
primary vortex; the dotted curve, to the secondary vortex.
(a) Voltage traces $V_p(t)$ and $V_s(t)$. (b) Schematic of time-dependent 
vortex position ($x$ is the x-coordinate of vortex
position; $a$ is the lattice constant.

\end{enumerate}

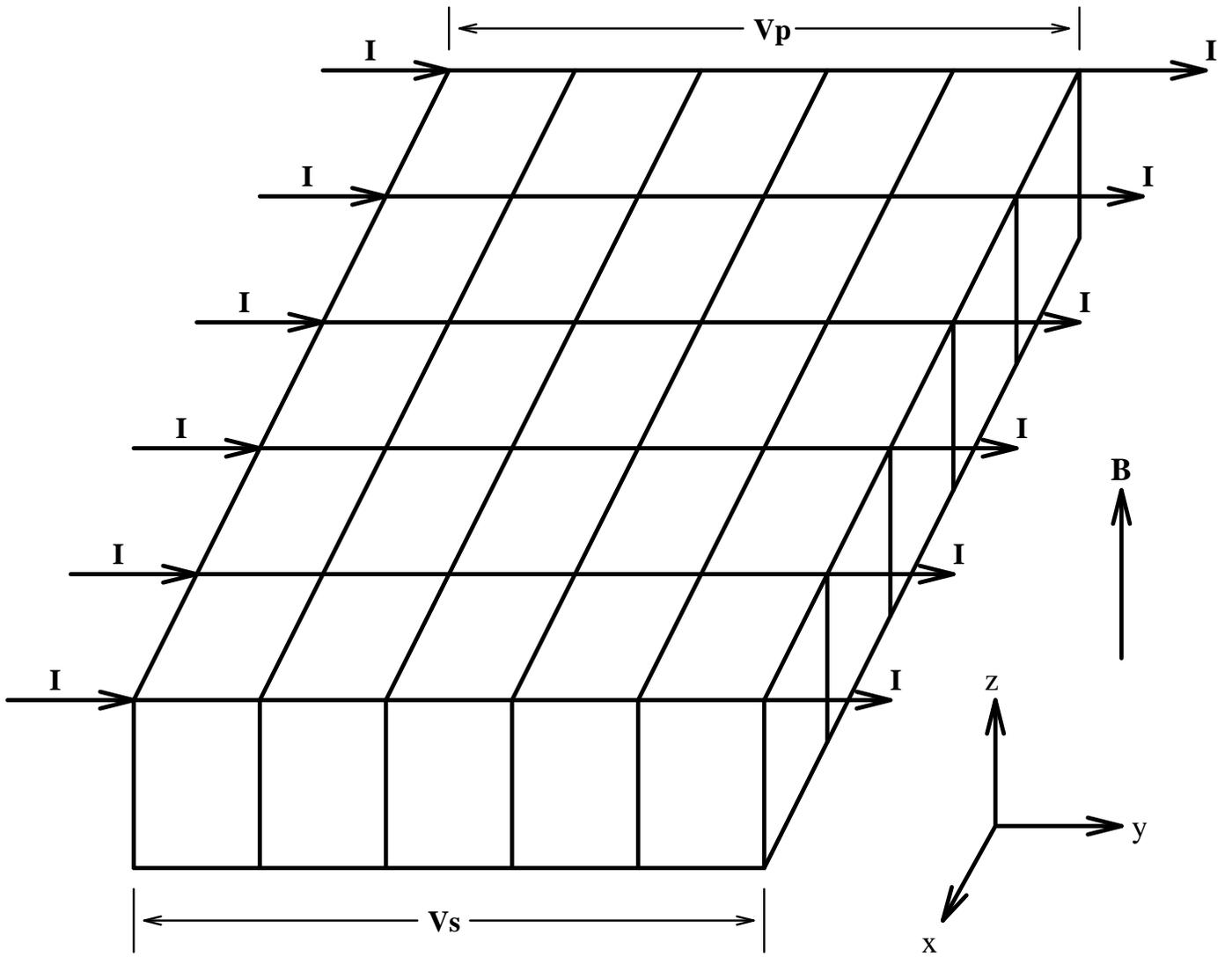
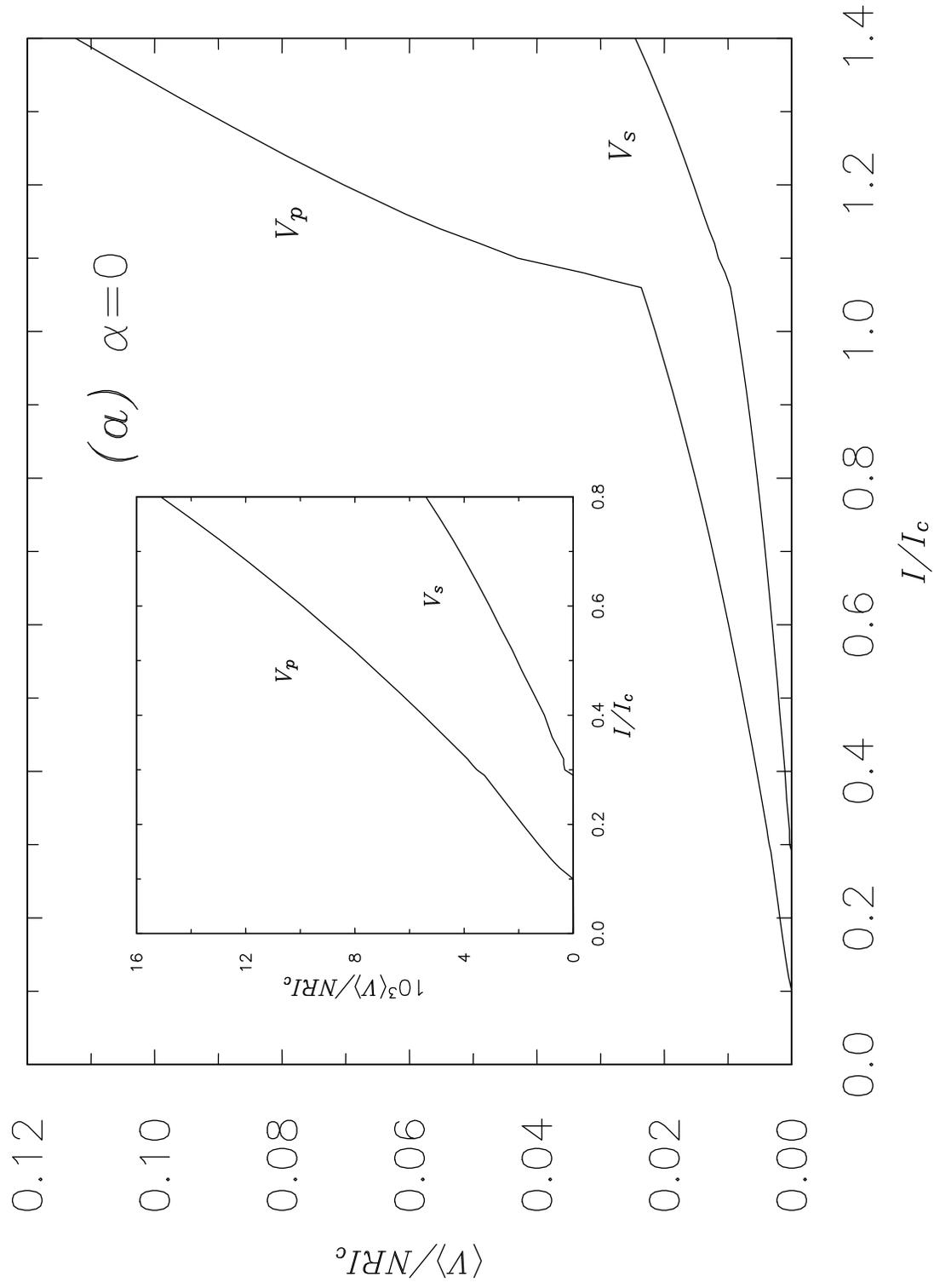
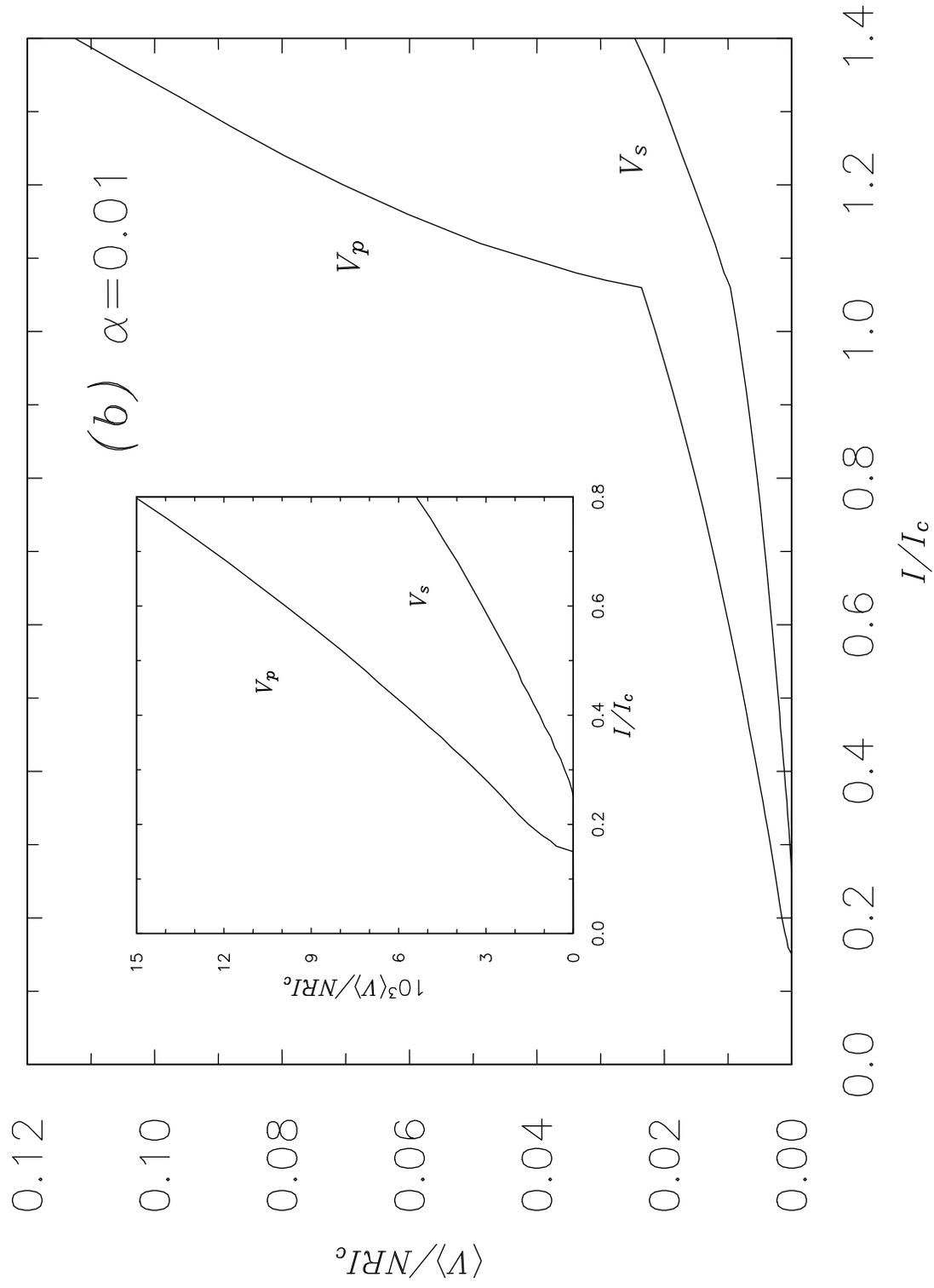
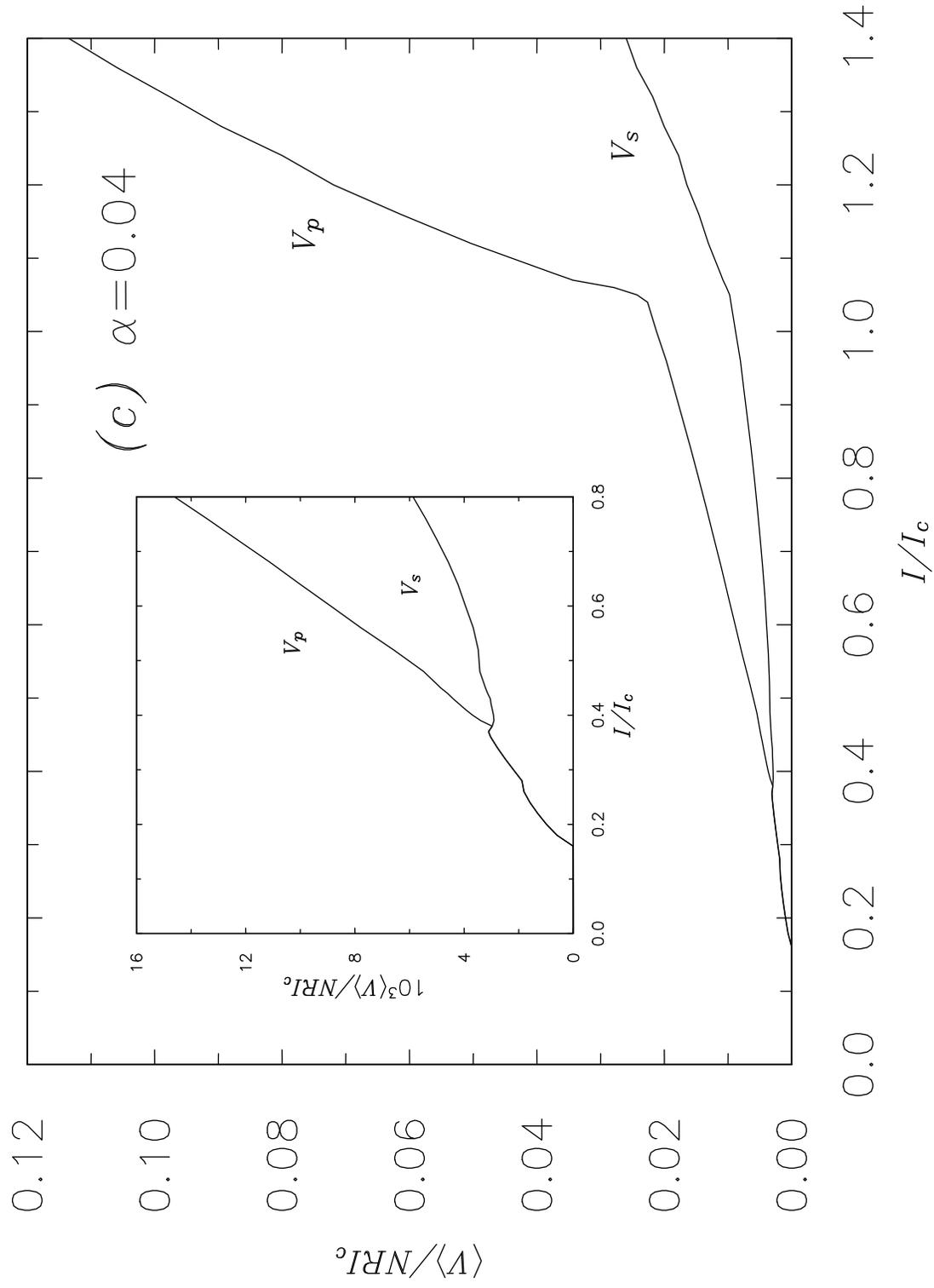
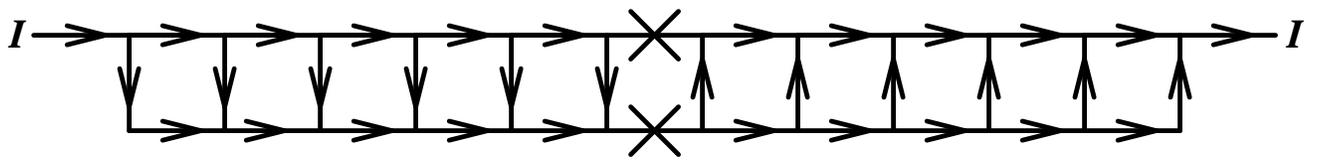
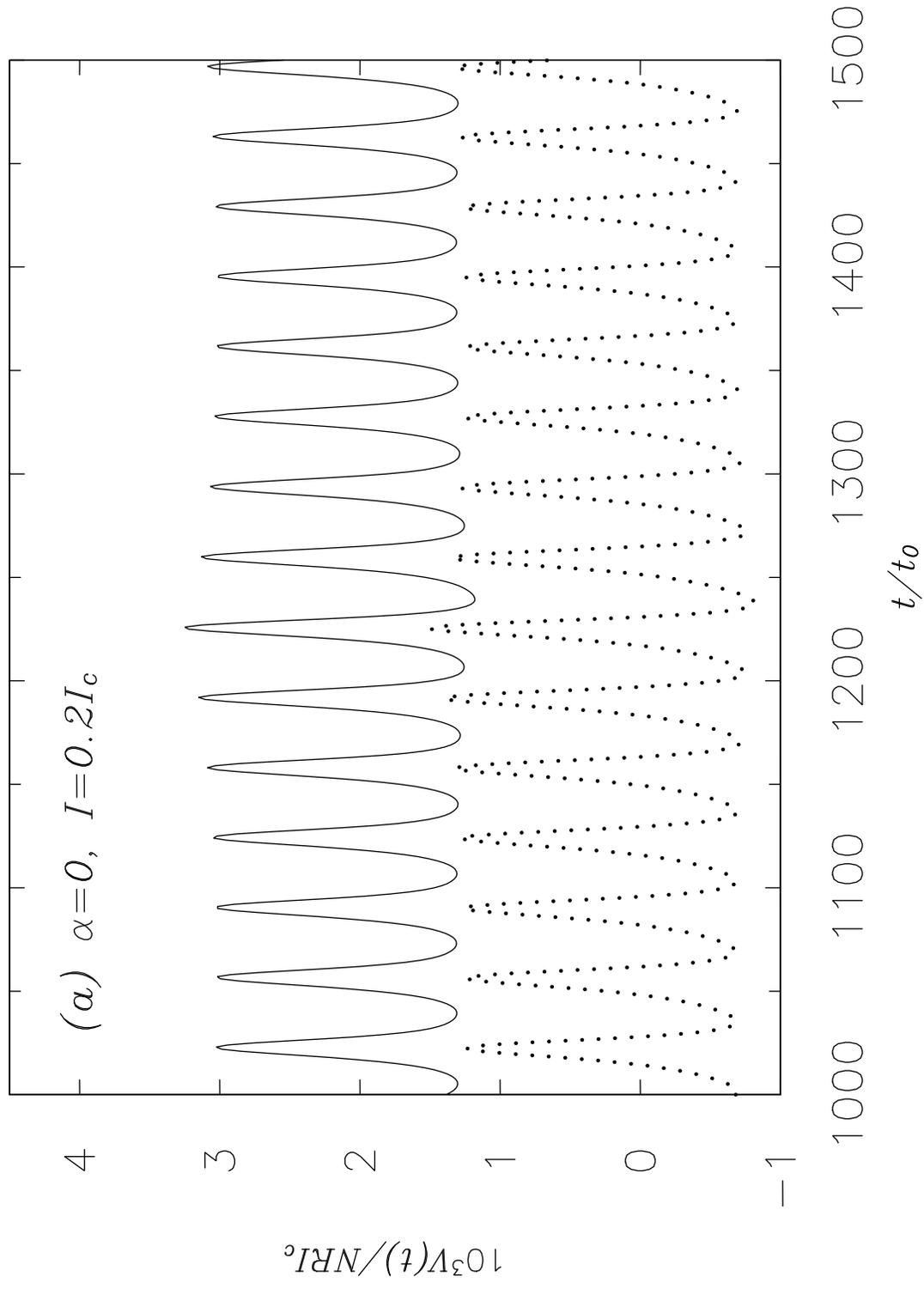
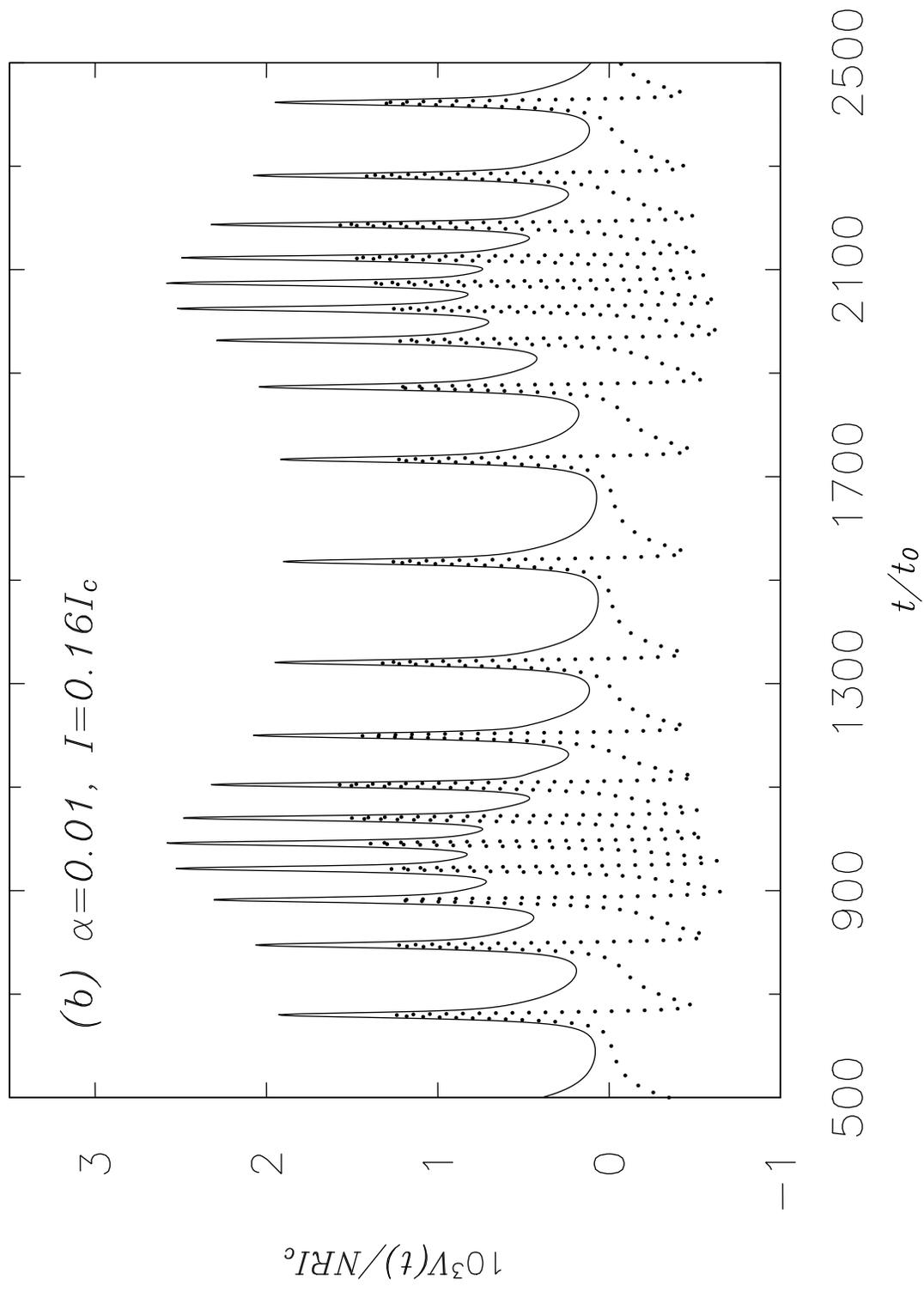
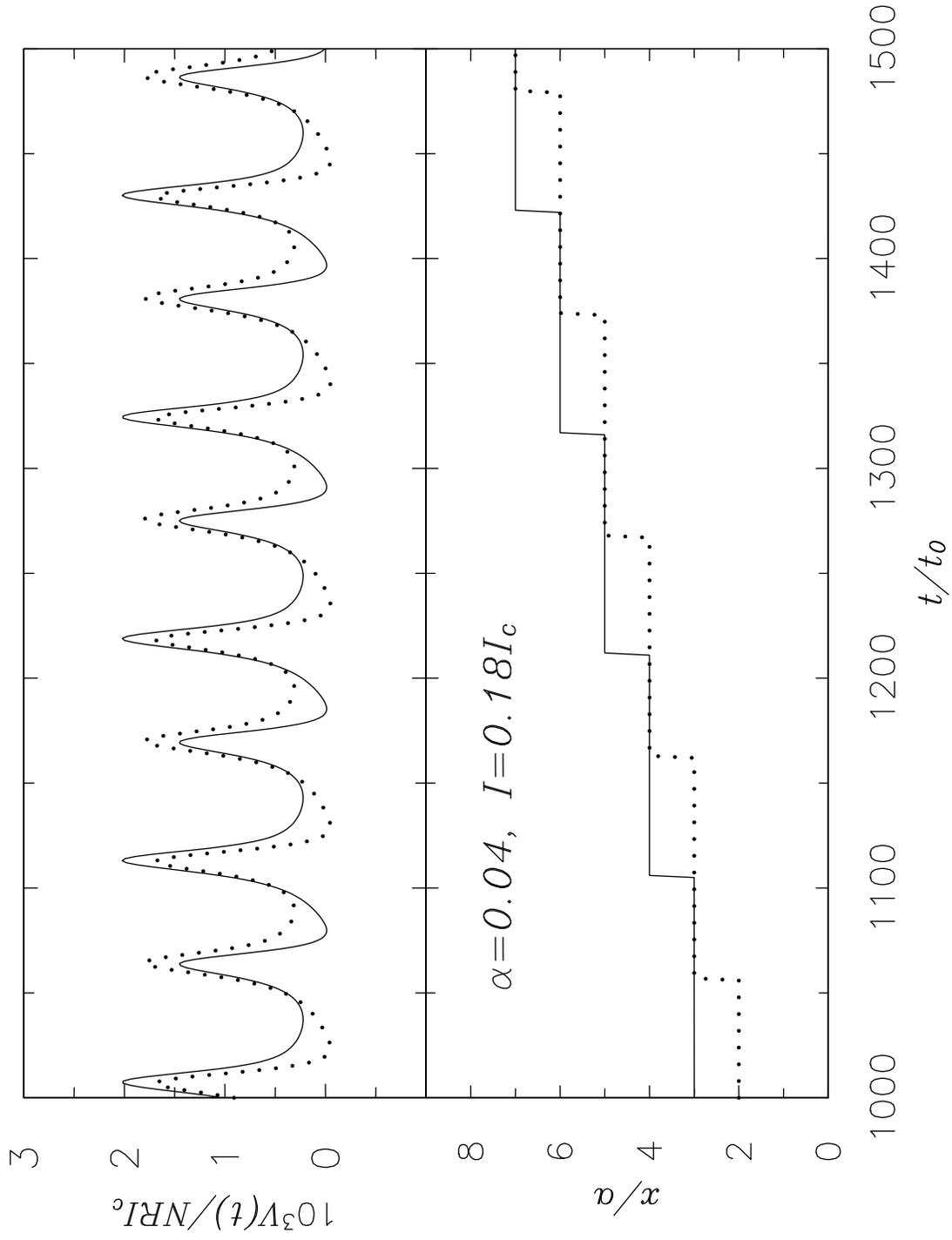
\end{document}